\documentclass[aps, prd, twocolumn, amsmath, floats,floatfix, superscriptaddress, nofootinbib]{revtex4}
\usepackage{amssymb,amsmath}
\usepackage{graphicx}
\usepackage{natbib}
\usepackage[OT1,T1]{fontenc}

\usepackage{soul}
\usepackage{color}

\begin{document}

\title{An estimate of the gravitational-wave background from the observed cosmological distribution of quasars}

\author{Nicolas Sanchis-Gual}
\affiliation{Departamento de Matem\'atica da Universidade de Aveiro and Centre for Research and Development  in Mathematics and Applications (CIDMA), Campus de Santiago, 3810-183 Aveiro, Portugal}

\author{Vicent Quilis}
\affiliation{Departament d'Astronomia i Astrof\'{\i}sica, Universitat de Val\`encia,
  c/ Dr. Moliner 50, 46100, Burjassot (Val\`encia), Spain}
\affiliation{Observatori Astron\`omic, Universitat de Val\`encia, c/ Catedr\'atico 
  Jos\'e Beltr\'an 2, 46980, Paterna (Val\`encia), Spain}
  
  \author{Jos\'e A. Font}
\affiliation{Departament d'Astronomia i Astrof\'{\i}sica, Universitat de Val\`encia,
  c/ Dr. Moliner 50, 46100, Burjassot (Val\`encia), Spain}
\affiliation{Observatori Astron\`omic, Universitat de Val\`encia, c/ Catedr\'atico 
  Jos\'e Beltr\'an 2, 46980, Paterna (Val\`encia), Spain}


\begin{abstract}
We study the gravitational-wave background from the observed cosmological quasar distribution. Using the DR9Q quasar catalogue from the ninth data release of the Sloan Digital Sky Survey (SDSS), we create a complete, statistically consistent sample of quasars from $z=0.3$ to $5.4$. Employing the spectroscopic information from the catalogue we estimate the masses of the supermassive black holes hosted by the quasars in the sample, resulting in a  log-normal distribution of mean $10^{8.32\pm0.33}M_{\odot}$. The computation of the individual gravitational-wave strains relies on specific functional forms derived from simulations of gravitational collapse and mergers of massive black hole binaries. The background gravitational-wave emission is assembled by adding up the individual signals from each quasar modelled as plane waves whose interference can be constructive or destructive depending on the quasar  evolutionary state. 
Our results indicate that the estimated gravitational-wave background discussed in this work could only be marginally detectable by LISA. This conclusion might change if more complete quasar catalogs than that provided by the SDSS were available.
\end{abstract}

\maketitle
\section{Introduction}
\label{section:intro}

Besides the extraordinary relevance of the landmark discoveries of Advanced LIGO and Advanced Virgo, the detection of gravitational waves from binary black hole (BBH) mergers and binary neutron stars (BNS) mergers~\citep{GW150914-prl, GW170814-prl,GWTC-1,GWTC-2,GW170817-prl} has updated previous estimates of the stochastic gravitational-wave backgrounds generated by these sources~\citep{GW170817-stochastic,GWO2-stochastic,BG}. Such backgrounds are a consequence of the superposition of the gravitational waves produced by a large  number of unresolved, independent sources, much too weak to be individually detected with current technology. In addition to those originating from BBH and BNS mergers, stochastic backgrounds of astrophysical origin exist for other types of sources, both transient, as supernovae or mergers of white dwarf binaries, and continuous, as spinning neutron stars or magnetars (see~\citep{Crocker:2017,Nelson:2019} and references therein). Moreover, gravitational-wave backgrounds in the nHz regime from supermassive black hole binaries (SMBHBs) have been extensively studied by the Pulsar Timing Array (PTA) 
community~\citep{sesana2008stochastic,sesana2010gravitational,cornish2013pulsar,sesana2015pulsar,IPTA1, NanoGrav1, IPTA2,NanoGrav2}.

In this work we estimate the relevance of the observed cosmological distribution of quasars from redshift $z = 0.3$ to 5.4 as a possible stochastic gravitational-wave background. The intense sustained activity of quasars during their $\sim 10^8$ yr lifetime (see~\cite{Schmidt:2017} for current observational estimates) is linked to violent processes associated with the presence of supermassive black holes (SMBHs) in the $10^6-10^{10} M_{\odot}$ mass range at their cores~\citep{Thorne:1976,Begelman:1980,Merritt:2005}. Proposed mechanisms for quasar formation typically involve the gravitational collapse of Population III supermassive stars (SMS) or other large distributions of gas clouds to yield massive black holes (MBH) that can subsequently merge to reach the observed final masses~\cite{Volonteri2010,bromm2020a,bromm2020b,bromm2021}.  The direct dynamical formation of MBH with $M\ge 10^6 M_{\odot}$ or the mergers of binaries of less massive such objects, may produce large-amplitude gravitational waves within reach of the LISA interferometer~\citep{bromm2003formation,amaro2012low,amaro2017laser}. Even if the association of SMBH birth from gravitational collapse cannot be connected with every instance of active quasars, such model allows to study the generation of an stochastic background by transient processes. However, no detectable gravitational radiation would be produced if SMBHs were instead the result of mass accretion onto significantly less massive seed black holes~\citep{begelman1979can, alexander2012, alexander2014rapid,mckinney2014three}.

Given the uncertainties in the actual processes operating during the formation of quasars, we assume in this work that the dynamics of the quasar phase is violent enough to produce low-frequency gravitational-wave bursts and ask ourselves if the combined signals from the observed quasar distribution might contribute to a new gravitational-wave background. For this purpose we use the DR9Q quasar catalogue from the Sloan Digital Sky Survey~\citep{ahn2012ninth} to establish a mock quasar distribution and compute the integrated amplitude of the gravitational-wave strain.  This computation relies on specific functional forms for the strain derived from simulations of both gravitational collapse~\citep{fryer2011gravitational} and massive BBH mergers~\citep{enoki2004gravitational}, which we use to produce the stochastic gravitational-wave background from the observed quasar sample.  We note that results on gravitational waves from catalogues have been reported before (e.g.~\cite{sesana2009gravitational,sesana2013systematic}) but those works have not  
considered the observed quasar distribution.

This paper is organized as follows. In Section~\ref{section:metho} we describe the methodology we follow to obtain our mock sample of quasars and the total gravitational-wave emission. Section~\ref{sec:results} presents our results for two different models. We close with Section~\ref{sec:conclusions} which presents our conclusions.

\section{Methodology}
\label{section:metho}

\subsection{Quasar sample}
\label{sec:sample}
\begin{figure}
\begin{center}
  \includegraphics[width=0.4\textwidth]{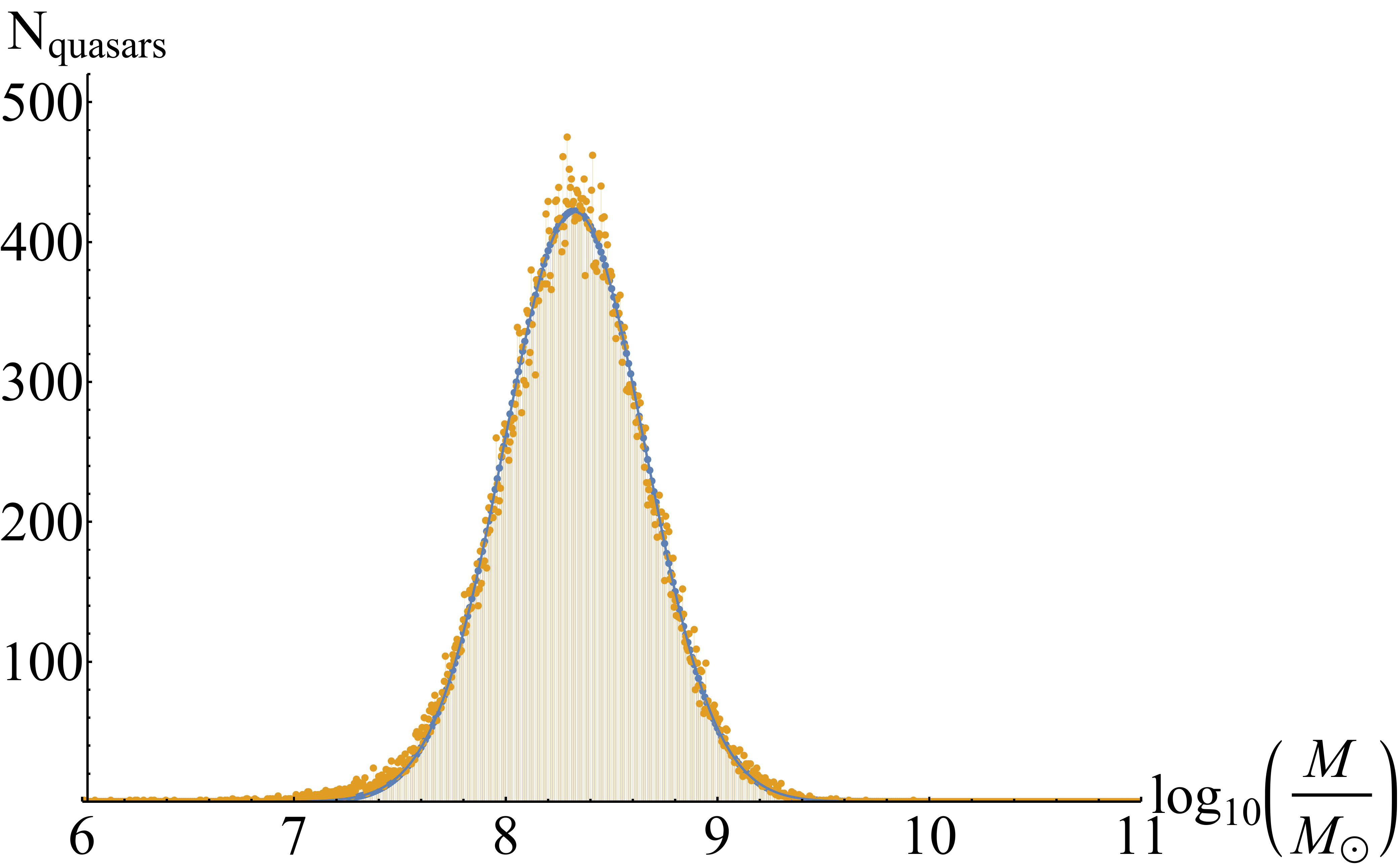}\\\hspace{-0.65cm}\includegraphics[width=0.4\textwidth]{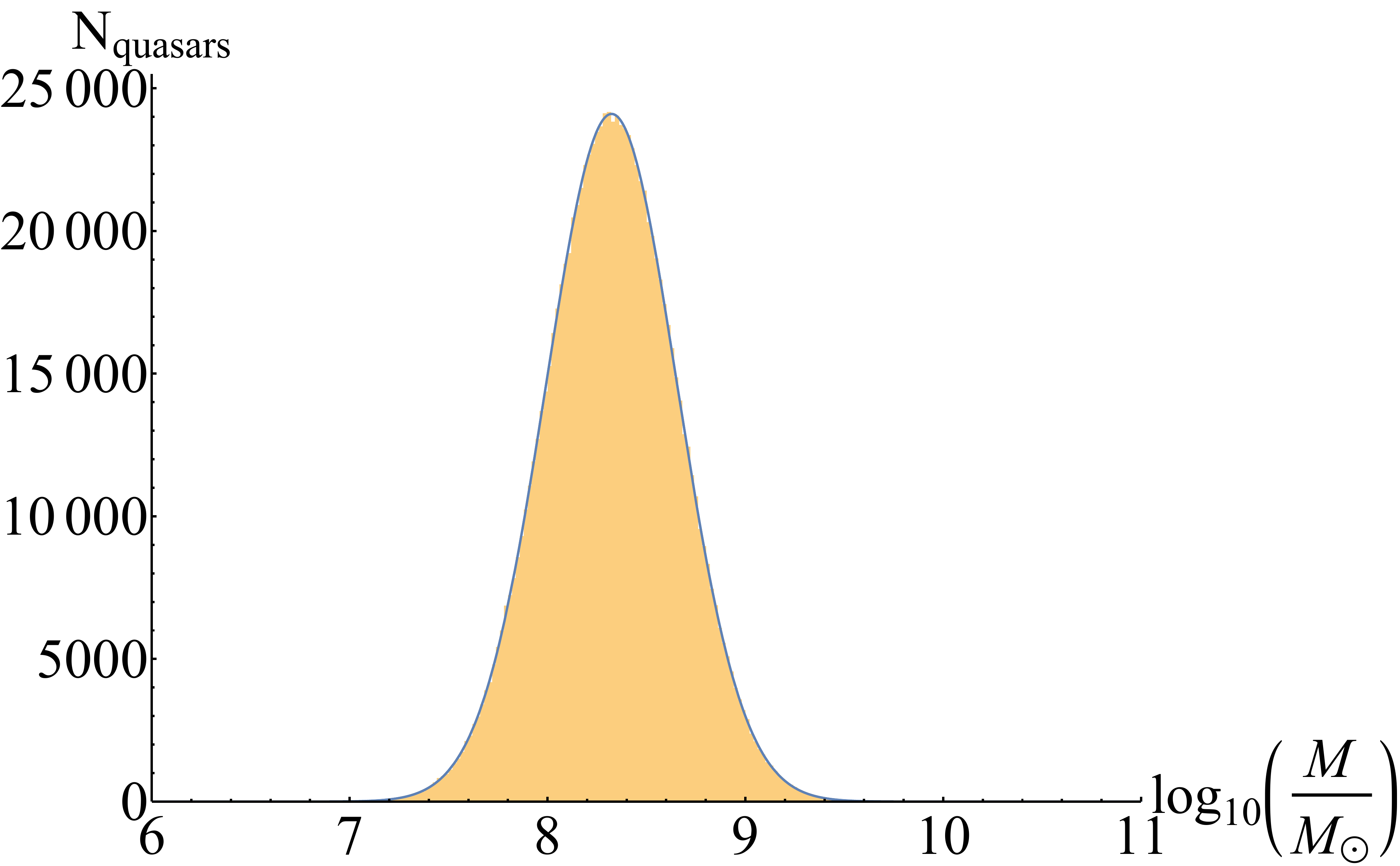} \includegraphics[width=0.4\textwidth]{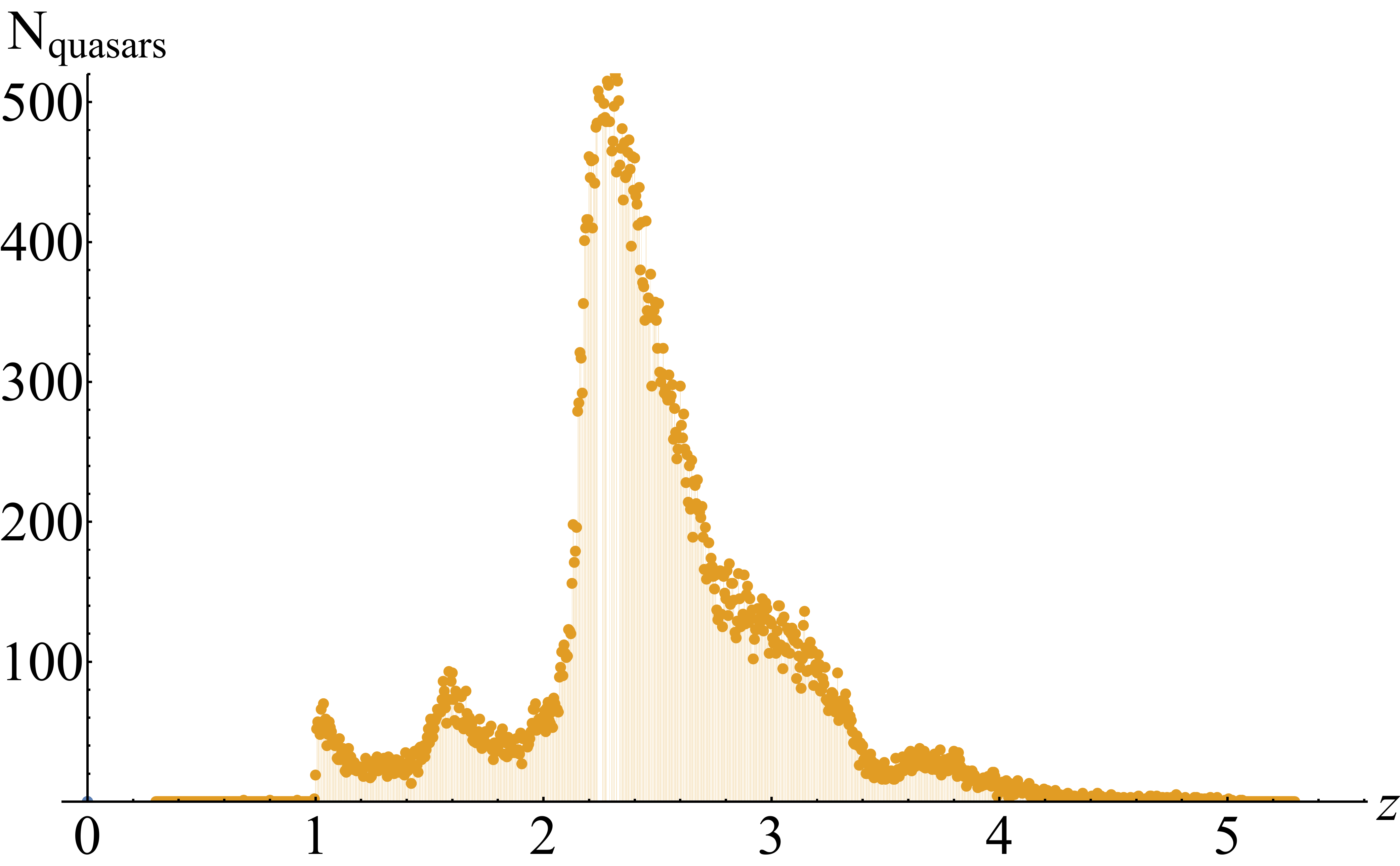}
\caption{Top panel: Histogram of the observed (DR9Q) SMBH mass. Middle panel: Histogram of the complete mock QSO catalogue. Bottom panel: Histogram of redshifts.}
\label{histogram}
\end{center}
\end{figure}

The ninth data release of the Sloan Digital Sky Survey~\citep{ahn2012ninth} includes a quasar catalogue (DR9Q)  with 87,822 objects. This catalogue covers several patches of the sky but is far from providing a full celestial sample and gives an average abundance of 27 QSO deg$^{-1}$. Despite the sparseness of the observed catalogue, a mock quasar catalogue covering the whole sky can be created. To do so, we first fit the masses of the SMBH hosted by the quasars to a log-normal distribution (top panel of Fig.~\ref{histogram}; the corresponding histogram of redshifts is plotted in the bottom panel). Quasars produce a characteristic electromagnetic emission due to accretion processes of matter onto their central SMBH. Their emission-line spectra present a great number of events in forbidden transitions. Such transitions include the Lyman alpha forest, but also C-III, C-IV, and Mg-II emission lines. The last two can be used to estimate the BH mass~\citep{Kong2006} by relating the BH mass to the emission-line bolometric luminosity, $L$,  and its fullwidth at half maximum of either transition,
\begin{eqnarray}
M_{\rm BH}({\text{C-IV}})&=&4.5\times10^{5}\,\biggl(\frac{L_{\text{C-IV}}}{10^{42}\,{\rm erg}\,{\rm s}^{-1}}\biggl)^{0.60\pm0.16}\nonumber\\
&&\times\,\biggl(\frac{\text{FWHM}_{\text{C-IV}}}{1000\,\rm{km}\,{\rm s}^{-1}}\biggl)^{2}\,\,M_{\odot}\label{MCIV}\,,\\
M_{\rm BH}(\text{Mg-II})&=&2.9\times10^{6}\,\biggl(\frac{L_{\text{Mg-II}}}{10^{42}\,{\rm erg}\,{\rm s}^{-1}}\biggl)^{0.57\pm0.12}\nonumber\\
&&\times\,\biggl(\frac{\text{FWHM}_{\text{Mg-II}}}{1000\,\rm{km}\,{\rm s}^{-1}}\biggl)^{2}\,\,M_{\odot}\,.\label{MMgII}
\end{eqnarray}

\begin{figure}
\begin{center}
\includegraphics[width=0.49\textwidth]{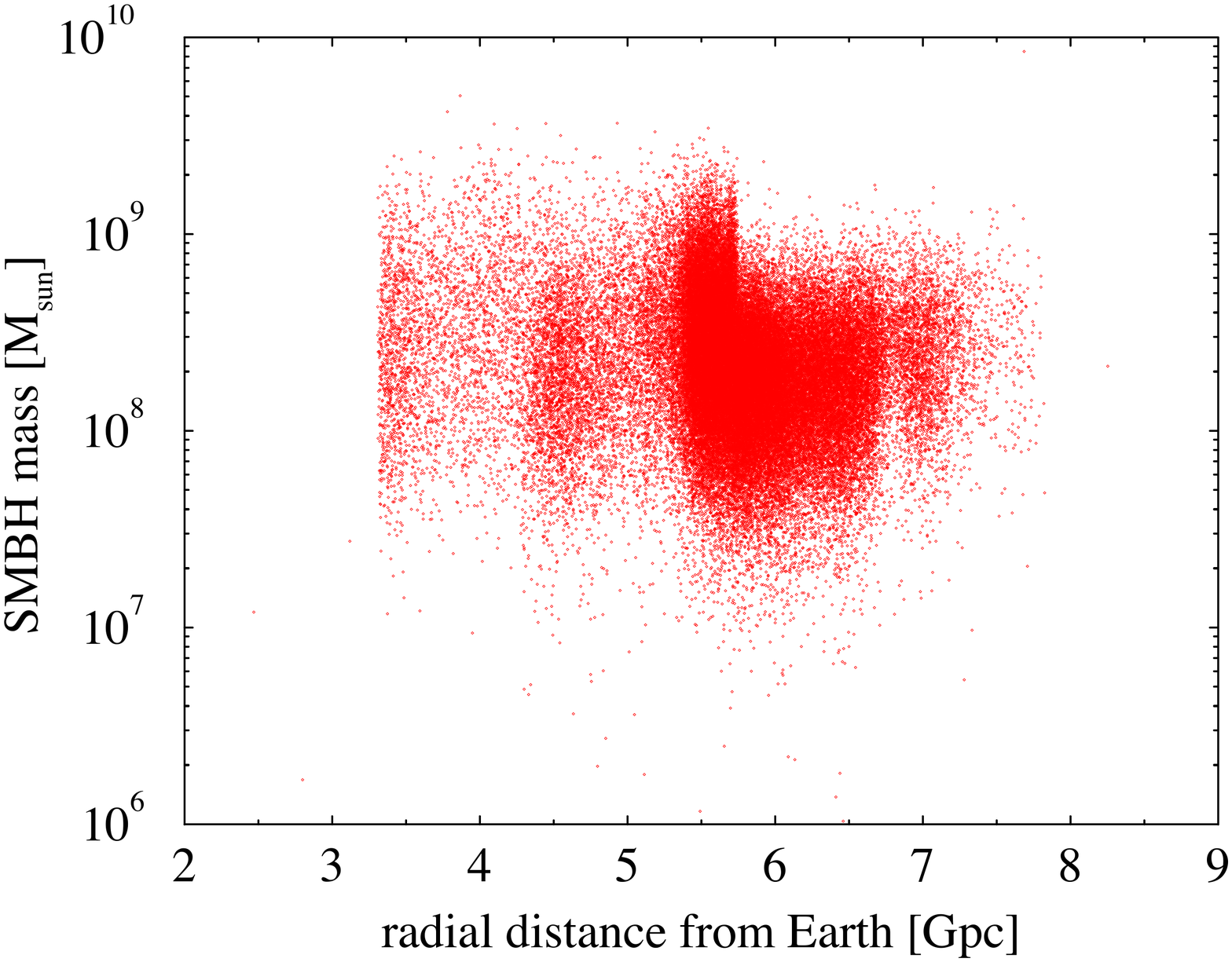}
\includegraphics[angle=90,width=0.49\textwidth]{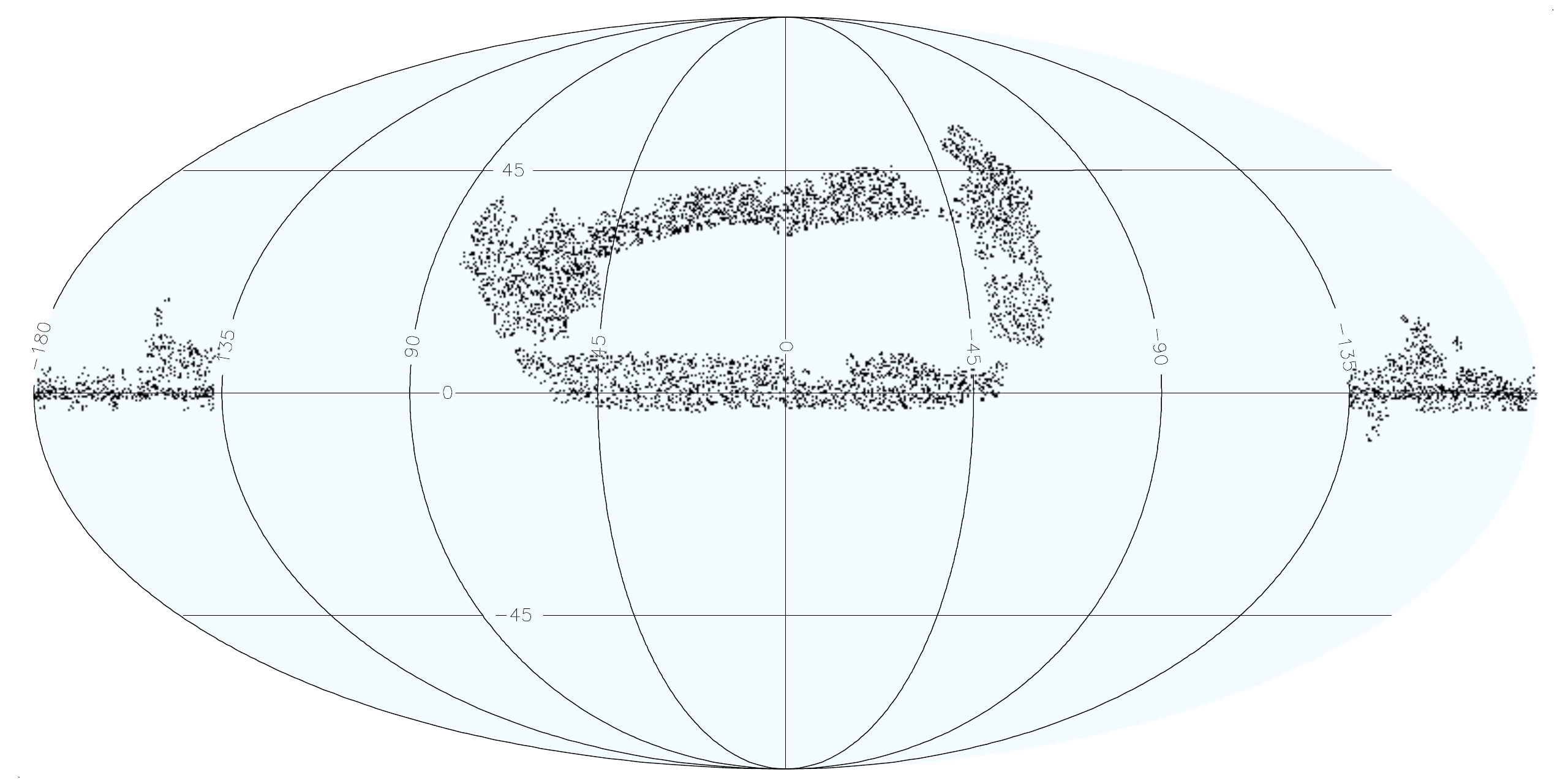}
\caption{Top panel: SMBH mass distribution of 81\% of the quasars from the SDSS III DR9Q catalog. Bottom panel: Sky positions of the 70,964 quasars whose SMBH mass was derived using C-IV and/or Mg-II emission line spectrometry.}
\label{fig:1}
\end{center}
\end{figure}

We estimate the masses of the BHs associated with the 87,822 quasars of the SDSS DR9Q catalog using Eqs.~(\ref{MCIV})-(\ref{MMgII}), where the luminosities are computed combining several spectrometric parameters provided by the catalogue for each quasar. When both emission lines are present we take the arithmetic mean of the two masses. Quasars are discarded if none of the lines are present. We are able to compute the BH mass for a total amount of 70,964 quasars, which represents 81\% of the total DR9Q catalog. BH masses span the interval between $10^{6} M_{\odot}$ and $10^{10} M_{\odot}$. In Fig.~\ref{fig:1} we plot the distribution of the SMBH masses as a function of the radial distance to Earth and the sky positions of the quasars to illustrate both properties of the sample. The resulting distribution of masses can be well described by a normal distribution with a mean value of  $M=10^{8.32 \pm 0.33}\,M_{\odot}$ where the error stands for $1\,\sigma$ deviation (see top panel of Fig.~\ref{histogram}). 

 Once we have identified the mass distribution of our sample of quasars, we proceed as follows:
 \begin{itemize}
 \item First, we produce a random sample of masses (see middle panel of Fig.~\ref{histogram}) following the redshift distribution of the real catalogue (see bottom panel of Fig.~\ref{histogram}). 
 \item Next, we distribute the quasars in three redshift bins.
 Using the \citet{landy1993bias} estimator, the correlation length and slope of the DR5Q catalog sample~\citep{schneider2007sloan} has been obtained in three studies, for  the low-redshift range $0.3\leq z\leq2.2$~\citep{ross2009clustering}, the intermediate-redshift range $2.2\leq z \leq2.8$~\citep{white2012clustering}, and the high-redshift range $2.9\leq z\leq5.4$~\citep{shen2007clustering}. For simplicity, we associate three different mean redshifts $z=$ $\lbrace 1.80, 2.44, 3.29\rbrace$ of the previous ranges. Following the redshift distribution shown in the bottom panel of Fig.~\ref{histogram} we distribute the proportional number of quasars to each redshift range, yielding $N^{z}_{\rm{quasars}}=\lbrace243081, 658064, 213854\rbrace$ quasars in each bin.
 \end{itemize}
 
 The final result is a sample containing $\sim 1,115\times10^6$ QSOs between $z=0.3$ to $5.4$ and spanning all $4\pi$ sr of the sky. This sample is statistically comparable to the observational catalogue in the sense that the two-point correlation function of both distributions is the same.  However, as the SDSS catalogue is not complete, the mock catalogue generated underestimates the total number of quasars. 

In order to fix the properties of each object in the new sample, we need to establish a cosmology. We adopt a $\Lambda$CDM model with cosmological parameters $H_{0} = 70$ km s$^{-1}$ Mpc$^{-1}$, $\Omega_{\rm m}= 0.3$, and $\Omega_{\Lambda} = 0.7$~\citep{spergel2007three}. In our approach we also need to know the luminosity distance, $d_{\rm L}$, of each quasar. For a given cosmology, this quantity is given by (see e.g.~\cite{kennefick2008infrared,Hogg}):  

\begin{eqnarray}\label{dL}
d_L = \frac{c(1+z)}{H_0} \int_0^z \frac{dx}{\sqrt{(1+x)^3 \Omega_m +
\Omega_\Lambda}}
\end{eqnarray}
where $c$ is the speed of light. 

\subsection{Gravitational-wave signal}
\label{sec:signal}

The crucial ansatz is this work is that  every observed quasar is emitting gravitational waves as a result of the violent processes occurring at their cores. Over cosmological time scales, the formation of quasars can be approximated as an instantaneous process. Therefore, the observation of a quasar by means of any electromagnetic signal directly implies that the emission of its gravitational radiation is also maximum, both in frequency and strain. Under this assumption, all quasars detected at a given redshift should be emitting the maximum of their gravitational-wave signals. Quasars may be active electromagnetically during different phases of AGN activity, e.g.~through accretion processes. However, these timescales are much longer than those of the gravitational-wave channel. We use this notion to underpin our simplified model in which the bulk of the gravitational-wave emission is either produced during the gravitational collapse of a SMS or in the late inspiral and merger of two MBHs. A more complex and realistic description should possibly include other AGN-related process and their gravitational-wave outputs besides the two main channels considered in this paper. Therefore, the results drawn from our models could be interpreted as a lower constraint to the total gravitational-wave background produced by the quasar distribution.

We assume that gravitational collapse is likely playing a major role on the dynamics and we model the gravitational-wave strain and frequency using results from the numerical relativity literature on direct gravitational collapse of supermassive stars~\citep{saijo2002collapse,saijo2004collapse,saijo-hawke,fryer2011gravitational,montero:2012,Shibata:2016}. Under this assumption, the dominant contribution to the gravitational-wave strain and frequency is given by the $l=m=2$ quasinormal-mode (QNM) ringdown of the SMBH
\begin{eqnarray}
f_{\text{QNM}}&=&2\times10^{-2}\,(1+z)^{-1}\,\biggl(\frac{10^{6}M_{\odot}}{M}\biggl)\,\text{[Hz]},\label{eq:frequency}\\
h_{\text{QNM}}&=&\epsilon\,6\times10^{-19}\biggl(\frac{\Delta E_{\text{GW}}/Mc^2}{10^{-4}}\biggl)^{1/2}\biggl(\frac{10^{-2}\,\text{[Hz]}}{0.5\,f_{\text{QNM}}}\biggl)^{1/2}\nonumber\\
&&\times\,\biggl(\frac{M}{10^{6}M_{\odot}}\biggl)^{1/2}\biggl(\frac{1\,\text{Gpc}}{d_{L}}\biggl),
\label{eq:strain}
\end{eqnarray}
where $\Delta E_{\text{GW}} \lesssim 10^{-4}Mc^2$ is the total radiated energy. We include the factor $\epsilon=0.1$ in Eq.~(\ref{eq:strain}), absent in the work of~\citet{saijo2004collapse}, to accommodate new results from recent simulations in which the amplitude is smaller~\citep{Shibata:2016,sun2017magnetorotational}. Note that, in this case, the emission of gravitational radiation is a short transient process (the collapse of SMS and SMBH birth) and cannot be associated with every stage of quasar activity. While limited by the duration of the burst, this scenario will show that we can model a stationary stochastic background from transient gravitational waves from quasars. This approach would be a simplification of a more realistic scenario in which the gravitational radiation from the evolution of SMS before undergoing gravitational collapse were considered.
In addition, we also consider the potential contribution of MBH mergers to the gravitational-wave signal~\citep{enoki2004gravitational}.  For a MBH binary, assuming that the orbit is quasi-stationary until the separation reaches the innermost stable circular orbit (ISCO), the gravitational-wave amplitude and maximum frequency at the ISCO are given by
\begin{eqnarray}
f_{\text{bin}}&=&4.4\times10^{-5}\,(1+z)^{-1}\,\biggl(1+\frac{M_{1}}{M_{2}}\biggl)^{1/2}\nonumber\\
&&\times\biggl(\frac{10^{8}M_{\odot}}{M_{1}}\biggl)\,\text{[Hz]},\label{eq:binfrequency}\\
h_{\text{bin}}&=&3.5\times10^{-17}\biggl(\frac{M_{\text{chirp}}}{10^{8}M_{\odot}}\biggl)^{5/3}\biggl(\frac{f(1+z)}{10^{-7}\,[\text{Hz}]}\biggl)^{2/3}\nonumber\\
&&\times\biggl(\frac{1\,\text{Gpc}}{d_{L}}\biggl)\,(1+z),
\label{eq:binstrain}
\end{eqnarray}
where $M_{\text{chirp}}=[M_{1}M_{2}(M_{1}+M_{2})^{-1/3}]^{3/5}$. For simplicity we consider one single merger of equal-mass components, namely $M_{1}=M_{2}=0.5\,M_{\text{BH}}$. This seems a conservative assumption considering that cosmological simulations show that SMBHs are formed by a combination of accretion and an important number of merger events involving 
similar objects~\citep{boothschaye2009}. The evolution
of the MBH orbit is not accounted for in our model, and the MBH gravitational-wave sources are always active during the simulated observation period. Therefore, we ignore the previous contribution of the earlier inspiral stages by taking the peak emission and the maximum frequency as an upper limit for the MBHs. We ignore the contribution of the earlier inspiral stages by taking the peak emission and the maximum frequency as an upper limit for the MBH mergers.

The gravitational-wave output from each quasar $k$ is modeled as a plane wave
\begin{equation}\label{eq:plane}
h_{k}(t) = h\,e^{i(\omega_{k} t + \phi_{k})},
\end{equation}
where $\omega_k$ is the frequency and $\phi_k$ is the phase. The frequency is assumed to be either complex, to account for the exponential damping of the QNM signal, or real, when dealing with the (continuous) inspiral signal of the MBH binary, respectively. The real part of $\omega$ is computed from Eqs.~(\ref{eq:frequency}) and (\ref{eq:binfrequency}) and the imaginary part is extracted from \cite{berti2006quasinormal}. Finally, the strain amplitude $h$ is evaluated from Eqs.~(\ref{eq:strain}) and (\ref{eq:binstrain}).

\section{Results}
\label{sec:results}

Considering the mock catalogue described in Section~\ref{sec:sample} and assigning a gravitational-wave signal to each object of this sample according to the method described in the previous section, the total gravitational-wave strain produced by the complete distribution of quasars is the sum of all interfering plane waves. In our study we compute the strain produced by two different models: QNM ringdown and binary BH merger. We assume that the emission from different quasars starts at different times, and the total strain reads:
\begin{equation}\label{eq:totalwave2}
h_{\text{T}}(t_{\rm{obs}}) = \sum^{N}_{k=1}\,H(t-t_{k})\,h_{k}(t_{\rm{obs}}-t_{k})\,,
\end{equation}
where $t_{\rm{obs}}$ is the observation time of the gravitational-wave detector, $N$ is the total number of quasars in our sample, $H(t-t_{k})$ is the unit step function and $t_{k}$ is the time at which the gravitational-wave signal from the $k$-th quasar is observed for the first time by the gravitational-wave detector. However, to avoid an unphysical non-stationarity in the binary model since the signals do not decay and the total strain would increase linearly with time, we compute the gravitational-wave strain only {\it after} all sources are emitting. In this case, we let the sources evolve during a year and then start the observation period.  

We further assume one year of observation time, $t_{\rm{obs}}=[0:32\times 10^{6}]$ s with a time resolution of $\Delta t_{\rm{min}}=100$ s. This choice is mainly computationally motivated since we need to resolve frequencies that span from $10^{-6}$ Hz to $10^{-4}$ Hz. Therefore, the computation requires high time resolution but also a long observation period. One year offers a good time window estimate for the frequencies in this range.

\begin{figure}[t!]
\begin{center}
\includegraphics[width=0.48\textwidth]{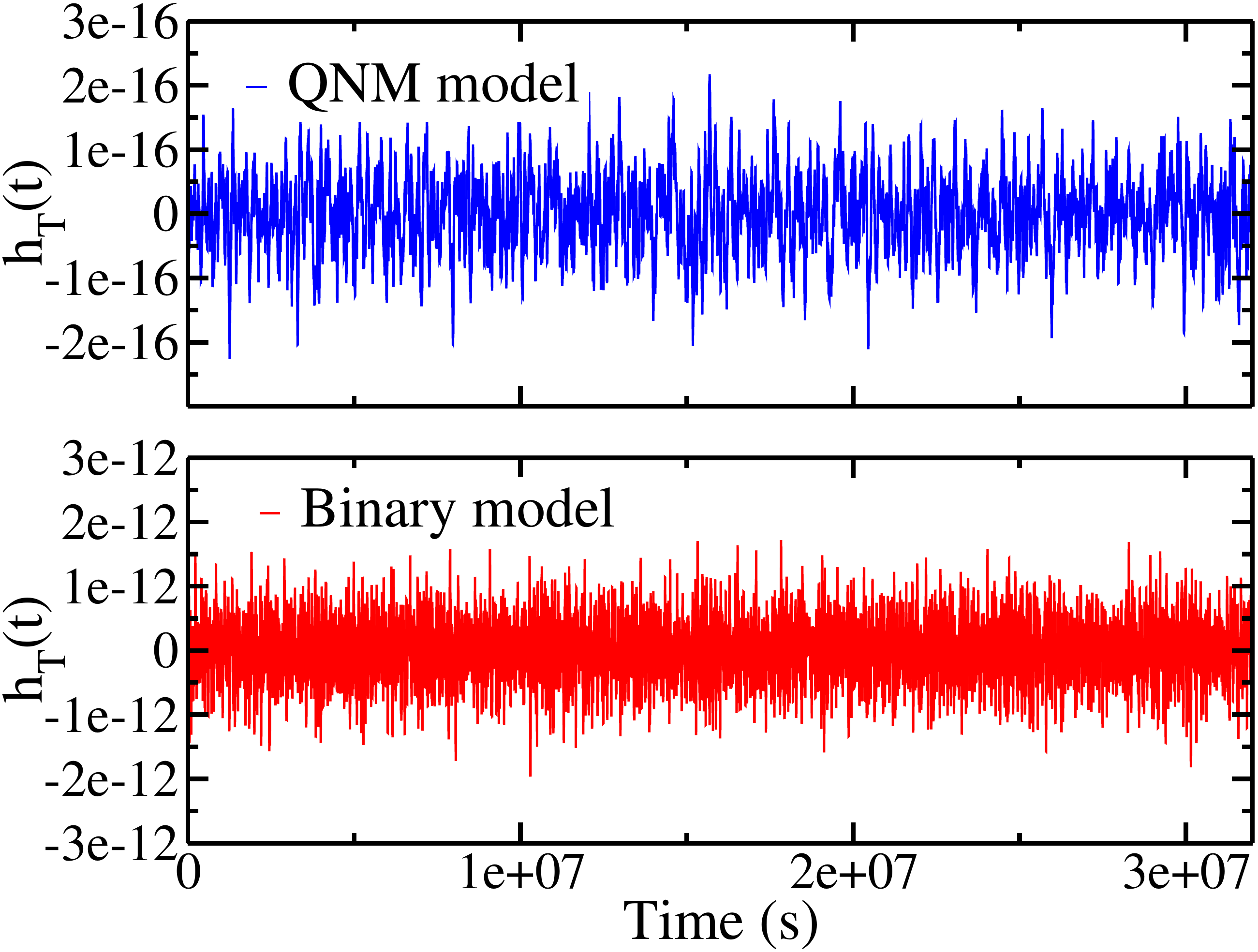}
\caption{Top panel: total gravitational-wave strain for quasar formation from direct gravitational collapse of SMS. Bottom panel: total gravitational-wave strain for quasar formation from MBH binary mergers after a year of evolution of the sources.}
\label{fig:GW-time}
\end{center}
\end{figure}

In Fig.~\ref{fig:GW-time} we plot the resulting gravitational waveform obtained from (\ref{eq:totalwave2}). This figure compares the gravitational-wave output associated with the two most favored mechanisms for quasar formation, namely gravitational collapse of SMS (blue curve) and mergers of MBH binaries (red curve). 

In the binary merger case we let the source evolve during a one-year span prior to the observation time. The amplitude grows with time in this phase because individual quasars start to emit gravitational waves at different times: their contributions add up to the total strain and the signals do not decay. At the beginning of the observation time all quasars are active and the amplitude is quasi-stationary (see bottom panel of Fig.~\ref{fig:GW-time}). On the other hand, since the amplitude of the QNM ringing signal decays exponentially with time, the emission is a short transient. However, each quasar is active at a different time. The top panel of Fig.~\ref{fig:GW-time} shows a total gravitational-wave strain with approximately constant amplitude from this type of transient signal.

Finally, Fig.~\ref{fig:GW-time} shows that the amplitude of the signal is about three orders of magnitude larger in the case of MBH mergers than for direct SMBH formation from the gravitational collapse of SMS. The frequency range is shown up to $\sim3\times10^{-4}$ Hz, which is the maximum frequency of our sample, obtained from the mass distribution and Eqs.~(\ref{eq:frequency}) and (\ref{eq:binfrequency}). Furthermore, the total amplitude in the binary case is larger because the gravitational wave of each quasar does not decay with time, since the frequency is real. In this case, a beating pattern in observed. The frequency distribution of our sample of quasars is peaked around a mean frequency of $10^{-4.78}$~Hz and a standard deviation of $\sigma=0.332$. Therefore, most of the quasars are emitting in a small frequency range, creating the interference phenomena. On the other hand, for the QNM model the beating is also present, with a mean frequency and the standard deviation, $10^{-4.57}$~Hz and $\sigma=0.330$.

\begin{figure}[t!]
\begin{center}
\includegraphics[width=0.48\textwidth]{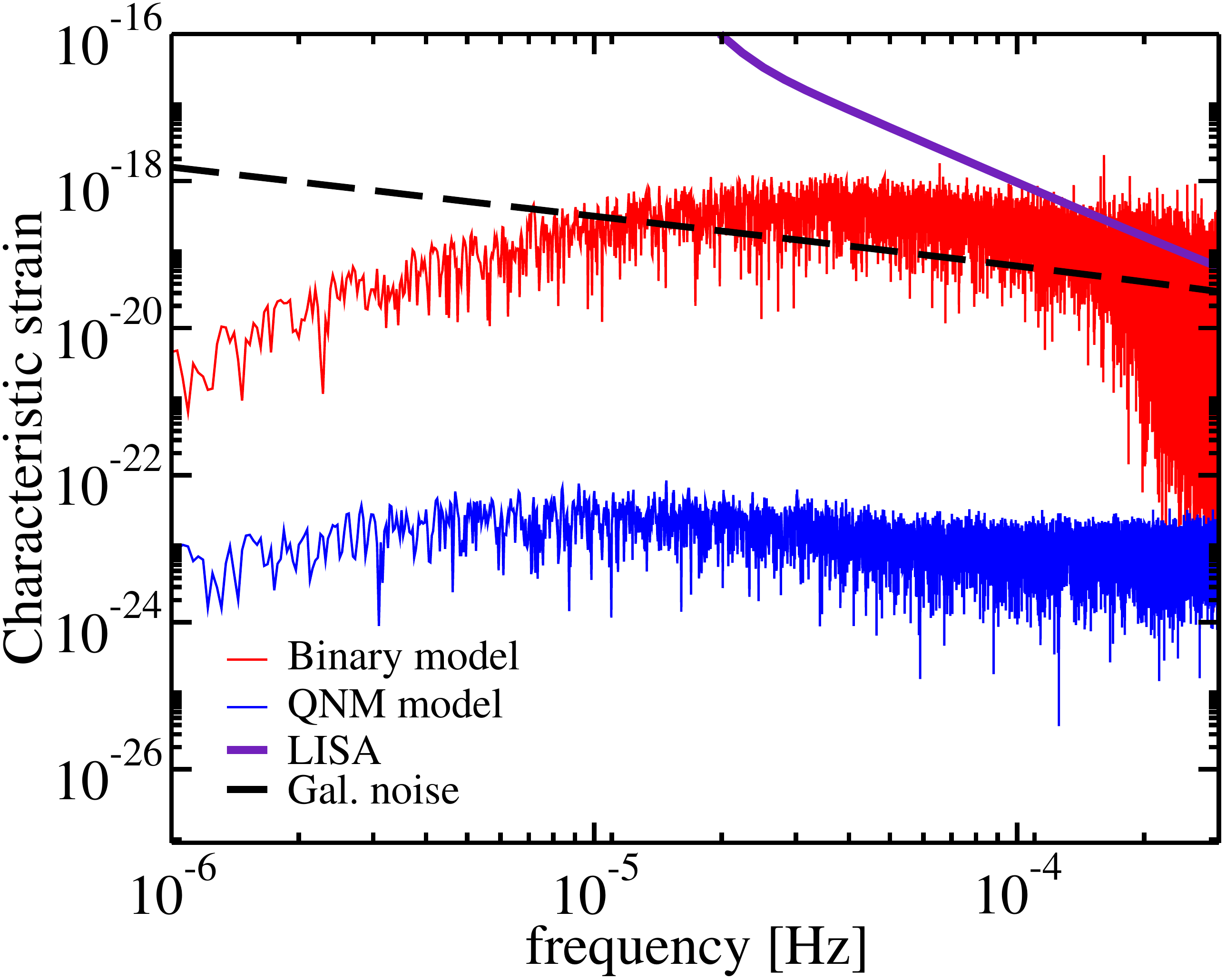}
\caption{PSD for the two mechanisms. The design sensitivity of LISA and the Galactic confusion noise component are shown for reference~\citep{babak2017science}.}
\label{fig:GW-PSD}
\end{center}
\end{figure}

By performing a Fourier transform of the waveforms we obtain the corresponding power spectra. Fig.~\ref{fig:GW-PSD} shows these spectra for the two mechanisms. This figure also includes the sensitivity curve of the LISA interferometer  and the Galactic confusion noise component, obtained from the analytic fits in~\cite{babak2017science}. Our results indicate that the amplitude of the estimated gravitational-wave background signal hardly falls within the sensitivity range of the LISA interferometer for the binary model but not for the QNM model. We note however that our estimation could be regarded as a lower bound given the fact that the SDSS catalogue is not complete.

\section{Conclusions}
\label{sec:conclusions}

In this work we have studied the gravitational-wave background produced by the observed cosmological quasar distribution. To do so we have built a mock sample of quasars covering the whole sky from the mass and redshift distributions of a real, but limited, sample extracted from the Sloan Digital Sky Survey DR9Q catalog~\citep{ahn2012ninth}. We have assumed that each quasar has emitted a gravitational-wave signal that reaches the Earth when a gravitational-wave detector is observing during a one-year period. Our approach has assumed that the SMBH at the center of a quasar has formed through two different mechanisms, (a) the gravitational collapse of a SMS and (b) the merger of two massive BHs. By computing the total contribution of all quasars from our mock sample we have provided estimates of the GW backgrounds produced in the two physical mechanisms investigated here.

While in this work we have only considered two possible quasar formation channels, this is still an open issue undergoing a deep debate. One possible manner to form SMBHs may be through multiple mergers of intermediate mass BHs with masses between 10 and 100 M$_{\odot}$ from Pop III stars (see for instance~\cite{marassi2011imprint}). However, this mechanism could be not efficient enough, failing to form the very SMBHs in quasars, as discussed by~\cite{djorgovski2008origins}. The other channel, commonly accepted to form the SMBHs present in quasars, is by mergers of black holes of similar masses. In this case, depending on the model, the number of mergers between black holes with masses larger than $10^6$M$_{\odot}$, would range between 1 and 10~\cite{sesana2009lisa}. Moreover, observational evidences seem to point that the presence of SMBHs in quasars could be well explained by a simple scenario of only a few galaxy mergers hosting SMBHs~\cite{treister2010major}. Other processes related to the evolution of AGN, such as accretion, could be considered as potential sources of gravitational waves. The inherent complexity of these process is out of the scope of the simple models employed in this paper, whose main goal has been to provide a lower estimate of total gravitational-wave signal produced by the quasar distribution.

As a consequence of our findings, the non-detection of the cosmological gravitational-wave background proposed in this work would also be a relevant result. Thus, the eventual absence of signal from the cosmological distribution of QSOs would imply that the process of formation and evolution of SMBHs within galaxies would be dominated by a quiescent form of accretion rather than by any violent phase of activity.

\acknowledgments
We thank the referee for his/her constructive criticism and useful suggestions and Julien Baur for his early work on this project during a research internship at the University of Valencia. Work supported by the Spanish Agencia Estatal de Investigaci\'on
(grants PID2019-107427GB-C33, PGC2018-095984-B-I00), by the Generalitat Valenciana (grant PROMETEO/2019/071), by the Center for Research and Development in Mathematics and Applications (CIDMA) through the Portuguese Foundation for Science and Technology (FCT - Funda\c{c}\~ao para a Ci\^encia e a Tecnologia), references UIDB/04106/2020 and UIDP/04106/2020 and by national funds (OE), through FCT, I.P., in the scope of the framework contract foreseen in the numbers 4, 5 and 6 of the article 23, of the Decree-Law 57/2016, of August 29, changed by Law 57/2017, of July 19. We acknowledge support  from the projects PTDC/FIS-OUT/28407/2017, CERN/FIS-PAR/0027/2019 and PTDC/FIS-AST/3041/2020. This work has further been supported by  the  European  Union's  Horizon  2020  research  and  innovation  (RISE) programme H2020-MSCA-RISE-2017 Grant No.~FunFiCO-777740. The authors would like to acknowledge networking support by the COST Action CA16104.

\end{document}